# Astro2020 Science White Paper

# Circumgalactic Gas and the Precipitation Limit

**Thematic Areas:**  ☐ Planetary Systems   ☐ Star and Planet Formation
  ☐ Formation and Evolution of Compact Objects   ☐ Cosmology and Fundamental Physics
  ☐ Stars and Stellar Evolution   ☐ Resolved Stellar Populations and their Environments
  ☒ Galaxy Evolution      ☐ Multi-Messenger Astronomy and Astrophysics

**Principal Author:**
Name:   G. Mark Voit
Institution:  Michigan State University
Email:   voit@pa.msu.edu
Phone:   517-884-5619

**Co-authors:** (names and institutions)
Arif Babul (Victoria), Iurii Babyk (Waterloo), Greg Bryan (Columbia), Hsiao-Wen Chen (Chicago), Megan Donahue (Michigan State), Drummond Fielding (Flatiron Institute) Massimo Gaspari (Princeton), Yuan Li (Berkeley), Michael McDonald (MIT), Brian O'Shea (Michigan State), Deovrat Prasad (Michigan State), Prateek Sharma (IISc Bangalore), Ming Sun (U. Alabama Huntsville), Grant Tremblay (CfA), Jessica Werk (Washington), Norbert Werner (MTA-Eotvos U./Masaryk U.), Fakhri Zahedy (U. Chicago)

**Abstract** (optional):
During the last decade, numerous and varied observations, along with increasingly sophisticated numerical simulations, have awakened astronomers to the central role the circumgalactic medium (CGM) plays in regulating galaxy evolution.  It contains the majority of the baryonic matter associated with a galaxy, along with most of the metals, and must continually replenish the star forming gas in galaxies that continue to sustain star formation.  And while the CGM is complex, containing gas ranging over orders of magnitude in temperature and density, a simple emergent property may be governing its structure and role.  Observations increasingly suggest that the ambient CGM pressure cannot exceed the limit at which cold clouds start to condense out and precipitate toward the center of the potential well.  If feedback fueled by those clouds then heats the CGM and causes it to expand, the pressure will drop and the "rain" will diminish. Such a feedback loop tends to suspend the CGM at the threshold pressure for precipitation.  The coming decade will offer many opportunities to test this potentially fundamental principle of galaxy evolution.

**Galaxy Evolution and Black-Hole Feedback**
Galaxy evolution depends critically on feedback. Without it, radiative cooling of cosmic gas would produce many more stars than are observed in today's galaxies [1-3]. The process of cooling, condensation, and accumulation of potentially star-forming gas into galaxies must therefore be self-limiting. In smaller galaxies, feedback from the supernovae that follow star formation can easily limit additional star formation because the specific supernova energy released by a typical stellar population greatly exceeds the specific binding energy of a galaxy's potential well. But supernova energy is insufficient to lift gas out of the potential wells of the most massive galaxies, implying that most of the feedback suppressing star formation in those systems comes from another source [4-9].

During the last decade, accretion of galactic gas onto supermassive black holes has emerged as the most likely candidate, and observations of the feedback mechanism have indicated deep connections between black-hole fueling and condensation of the circumgalactic medium (CGM). The best places to observe black-hole feedback in action have been the centers of galaxy clusters, because the ambient CGM in those systems glows brightly in X-rays. There, the central black hole blows bubbles of radio-emitting relativistic plasma into the surrounding medium, producing weak shocks, turbulence, mixing, and cosmic rays [10-15]. Collectively, those heating channels introduce enough energy to compensate for radiative cooling of the ambient medium, thereby limiting star formation. However, the microphysical processes that distribute and thermalize the feedback energy remain uncertain and a subject of intense theoretical interest (see white paper led by Ruszkowski).

Remarkably, observations show that this black-hole feedback mechanism is finely tuned, despite its chaotic appearance. Strong radio emission and circumgalactic bubbles appear almost exclusively in the subset of cluster cores with a hot-gas cooling time < 1 Gyr at ~10 kpc from the center [16,17]. Also, the correlation between a short central cooling time and the presence of multiphase gas, including H$\alpha$ nebulosity, molecular clouds, and star formation, is nearly one-to-one [17-19]. In other words, the central engine supplies enough energy to compensate for cooling in systems that currently require it and is far less active in systems that do not currently need feedback.

The introduction of feedback energy is also surprisingly gentle. Naively, one might expect specific entropy levels in the ambient medium to be largest near the central heat source, but in fact the opposite is true [20-22]. Cluster cores with active central engines tend to have low central entropy and entropy gradients that rise with radius, indicating that their energy input is thermalized over a large volume extending tens of kiloparsecs into the surrounding medium, allowing the ambient gas to remain stratified and close to hydrostatic equilibrium.

**The Precipitation Limit**
So how does the accretion engine on sub-parsec scales manage to adjust itself so nimbly to the thermodynamic properties of the ambient medium on scales exceeding $10^5$ parsecs? One promising hypothesis relies on a phenomenon sometimes called "precipitation" [23]. Closely related and conceptually overlapping variants go by the names "cold feedback" [24], "chaotic cold accretion" [25], and "stimulated feedback" [26]. All are inspired by the observed correlations between short central cooling time, strong radio power, and the presence of multiphase gas, which suggest that feedback is strongly boosted as the ambient medium transitions to a multiphase state. While the ambient medium is homogeneous, its accretion rate remains similar to the classic Bondi rate, which is limited by the specific entropy of the gas. But if the ambient medium can "precipitate" by condensing into cooler clouds that rain toward the center, the accretion rate can rise by orders of magnitude [24,25,27-29]. In order to fuel the black hole, at least some of those falling clouds must have low specific angular momentum, and that can be achieved in a turbulent medium with a velocity dispersion sufficient to populate the low end of the angular momentum distribution with cold clouds. If that condition is satisfied, then a "chaotic cold accretion" process will rapidly channel fuel toward the central engine [25,30-32].

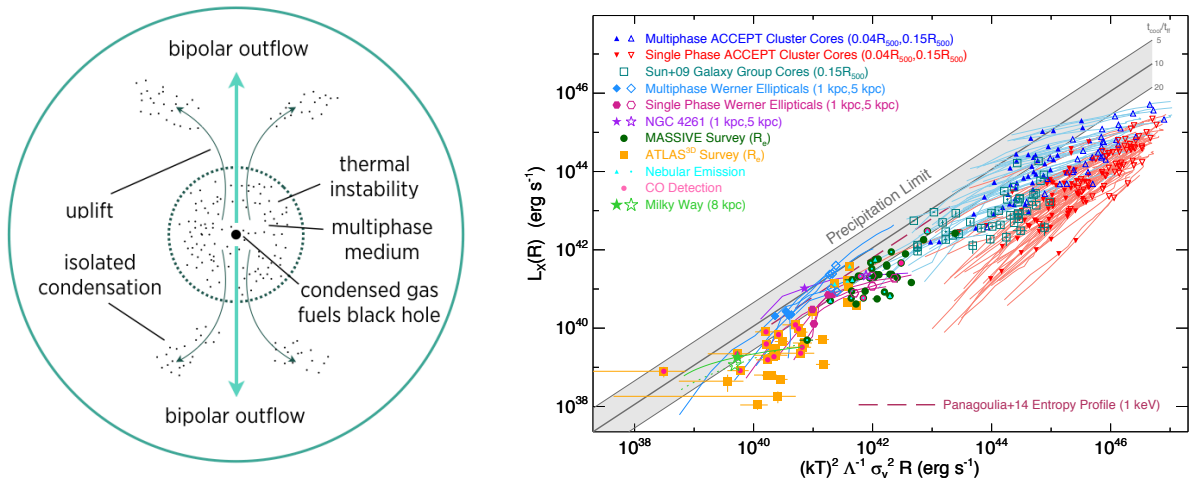

*Figure 1:* (Left) Schematic representation of precipitation-regulated feedback in a massive galaxy. (Right) Evidence for a pervasive precipitation limit on circumgalactic electron density from observations of X-ray luminosity ($L_X$) within a given radius (R) around galaxies more massive than the Milky Way [36].

Observational evidence supporting the precipitation hypothesis comes primarily from measurements of the ratio of gas cooling time ($t_{cool}$) to free-fall time ($t_{ff}$) in the ambient media around massive galaxies [23,33-36]. Numerous simulations of precipitation-regulated feedback show that it tends to suspend the ambient medium in a state with $10 < \min(t_{cool}/t_{ff}) < 20$ [25,37-41]. Conventional wisdom long held that an entropy-stratified medium would transition from a homogenous state to a multi-phase state when $t_{cool}/t_{ff}$ dropped below unity, but that criterion applies only to a nearly static medium. The critical value rises by roughly an order of magnitude if either turbulence or bulk uplift counteracts the buoyant restoring forces that would otherwise suppress incipient gas condensation and formation of a multiphase medium [42-44].

There is not yet a complete theory of precipitation that derives the critical value of $t_{cool}/t_{ff}$ from first principles, and many of the details remain subjects of active debate. For example, the stimulated feedback model posits that ~1 keV gas cools and condenses as it rises in the wakes of buoyant bubbles but does not necessarily fuel the central engine [26]. Helpfully, cosmological numerical simulations are now beginning to resolve the condensation and feedback processes resulting from uplift and turbulence [45] and will inform observational tests of the various possibilities using both X-ray observatories and high-resolution CO observations with ALMA [46].

**Precipitation in Smaller Galaxies**
The simplicity and generality of the physical principles that appear to produce a precipitation limit in galaxy-cluster cores suggest that the precipitation limit might apply more broadly, perhaps even to galaxies regulated primarily by supernova feedback. At its most general, the precipitation hypothesis simply states that ambient circumgalactic gas with $t_{\rm cool}/t_{\rm ff} < 10$ cannot persist without producing a rain of cold clouds that fuel feedback, thereby heating the CGM, raising $t_{\rm cool}$, and diminishing the precipitation. Galactic systems with this feature inevitably regulate themselves so that their gaseous halos remain near the threshold for precipitation. And if this principle applies to galaxies of all masses, then it may be at the root of many other patterns observed among galaxies, including the mass-metallicity relation, the relationship between stellar mass and circular velocity, and the correlation between stellar velocity dispersion and central black-hole mass [47].

Observations are just beginning to probe whether galaxies less massive than those in cluster cores adhere to the precipitation limit. Some examples are shown in Figure 2, which illustrates electron density as a function of $r/r_{500}$, where $r_{500}$ is the radius encompassing a mean mass density 500 times the cosmological critical density. Panel (a) shows the galaxy-cluster observations that first drew attention to the precipitation limit. Panel (b) shows how the precipitation limit relates to observations of $n_e(r)$ around early-type galaxies in halos of mass $\sim 10^{13}$ $M_{\rm Sun}$. Panel (c) shows a recent compilation of constraints on $n_e(r)$ around the Milky Way. Panels (d), (e), and (f) show ambient CGM densities inferred from absorption-line observations of photoionized clouds around galaxies with stellar masses in the range $10^{9.5}$ to $10^{11.5}$ $M_{\rm Sun}$. Collectively, they indicate that the precipitation limit may be a general feature of the CGM on all mass scales, and this intriguing finding stands as an important benchmark for future simulations of feedback, suggesting that regulation via precipitation ought to emerge naturally as those simulations achieve greater realism.

**Precipitation Predictions for the CGM**
During the coming decade, observations of the CGM around lower-mass galaxies will be critical for furthering our understanding of how feedback regulates galaxy evolution through its impact on the CGM (see the white papers led by Chen, Oppenheimer, and Peeples). Several predictions of precipitation-limited models will be testable, given the proper resources:

- *Limiting Pressure.* The strongest prediction is that feedback should place an upper limit on ambient CGM pressure corresponding to $t_{\rm cool}/t_{\rm ff} \sim 10$ across the entire mass range of galaxies. It can be tested by seeking correlations between Sunyaev-Zeldovich CMB distortions, X-ray luminosity, and galaxy mass [48,49] as well as pressure measurements derived from UV observations of photoionized CGM clouds [50].
- *Kinematics.* Ambient gas in the CGM of a precipitation-regulated galaxy is expected to be circulating at sub-Keplerian speeds, with a 1D velocity dispersion roughly half that of the central galaxy's stars. High-resolution UV, optical, and radio spectroscopy will be indispensable for mapping the CGM velocity field and deconvolving multiphase structures projected along the line of sight, while X-ray spectroscopy with XRISM [51], and eventually *Athena,* will radically advance our understanding of the hot-gas kinematics.

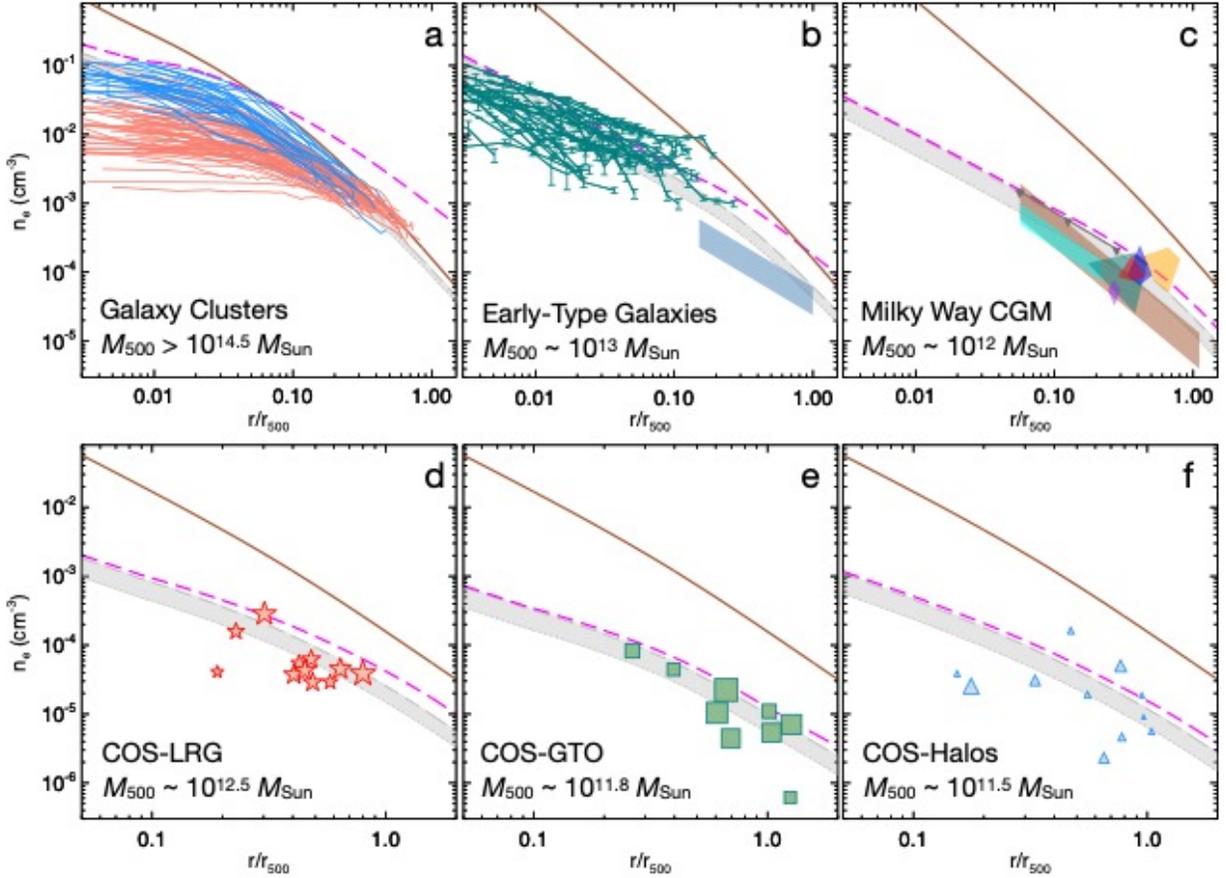

***Figure 2:*** *Measurements of electron density in the ambient CGM as a function of radius in units of the cosmological scale radius $r_{500}$. In each panel, a solid brown line shows the expected electron-density profile without feedback, and a dashed magenta line shows the precipitation limit corresponding to $t_{cool}/t_{ff} = 10$ for a representative potential well in the mass range of that population. Grey regions show hybrid profiles combining the cosmological and precipitation limits for $10 < t_{cool}/t_{ff} < 20$ [52].*

(a) *Galaxy clusters from the ACCEPT sample [21]. Clusters with multiphase cores are in blue and those with homogenous cores are in red.*

(b) *Early-type galaxies with $\sigma_v \sim 300$ km/s. Green lines show electron-density profiles derived from X-ray observations [56], and the blue strip shows constraints derived by [48] from X-ray and CMB stacks of luminous red galaxies. (Profiles that extend above the dashed magenta line do not necessarily represent violations of the precipitation limit because those galaxies may be in potential wells deeper than indicated by the galaxy's central stellar velocity dispersion.)*

(c) *An assortment of constraints on the Milky Way's CGM [52].*

(d) (e) (f) *Ambient CGM densities inferred from absorption-line observations of photoionized clouds around galaxies of differing stellar mass [50].*

- **Thermal Structure.** Precipitation-limited CGM models make definite predictions for X-ray absorption-line column densities that depend surprisingly weakly on CGM metallicity [52]. These predictions will eventually be testable with *Athena* but could be tested sooner with a spectroscopic X-ray mission similar to ARCUS [54]. Precipitation-limited models also predict that vertical uplift of the ambient medium should result in adiabatic cooling capable of boosting the CGM column densities in lines such as Ne VIII, O VI, and N V, which require UV spectroscopy [52].
- **Stimulated Condensation.** In the precipitation framework, the strong molecular outflows observed in some central cluster galaxies are thought to result from particularly strong updrafts that stimulate condensation of ambient gas into dense molecular clouds [26,42,53]. If that interpretation is correct, then detailed observations of molecular outflows in smaller galaxies should reveal that they are precipitating and do not necessarily require the gas to remain in molecular form as it accelerates to escape speed from the galaxy.

Furthermore, the precipitation framework predicts that star-formation quenching should result naturally from accumulation of stellar mass around the central black hole. As the galaxy's stellar potential well deepens, the pressure gradient required to drive galactic gas out of it becomes steeper. If an increase in the entropy gradient accompanies that increase in the pressure gradient, then buoyant suppression of condensation will make the CGM increasingly resistant to condensation and production of multiphase gas [42]. In other words, there should be a qualitative change in the galaxy's response to feedback as the central stellar velocity dispersion rises and forces condensation to occur centrally, in the vicinity of the supermassive black hole. Observing this phenomenon in a representative sample of galaxies will require high-resolution X-ray imaging, possible with *Lynx* but not *Athena*.

**Other Applications of the Precipitation Limit**
While the jury is still out on whether the current theoretical interpretation of the precipitation limit is correct, it still has considerable pragmatic value as a fitting formula for representing the CGM properties of galactic systems with halo masses of $10^{12}$ to $10^{14}$ $M_{\mathrm{Sun}}$. Specification of CGM properties in that mass range will be necessary to maximize the scientific return on upcoming efforts like *Euclid*, which aspire to measure cosmic structure formation with unprecedented accuracy using the weak-lensing power spectrum [51]. Because of feedback, baryonic mass does not trace dark matter on scales of 10 to 100 kpc, by an amount depending on halo mass, with observable consequences for interpretations of cosmic shear. Similarly, the next generation of microwave background experiments aspire to measure structure imprinted on the CMB by the Sunyaev-Zeldovich effect, in both thermal and kinetic forms [48]. Interpretation of those measurements will require an accurate and physically motivated parametric model for the circumgalactic pressure distribution at 10 to 100 kpc, and the precipitation limit provides one that is currently consistent with a broad range of CGM observations.

The precipitation limit also serves as an important guideline for planning of future X-ray imaging missions, such as *Lynx*. Given the observations in Figure 2, it appears likely that low-redshift galactic systems rarely exceed the density threshold at which $t_{\mathrm{cool}}/t_{\mathrm{ff}} = 10$, but systems with $10 < \min(t_{\mathrm{cool}}/t_{\mathrm{ff}}) < 20$ ought to be common. A mission capable of imaging them would yield unique and powerful insights into the feedback processes that govern galaxy evolution.